# Hydrogenation Properties of the TiB$_x$ Structures

R. Žitko[1], H. J. P. Van Midden[1], E. Zupanič[1], A. Prodan[1*], S. S. Makridis[2,5], D. Niarchos[3] and A. K. Stubos[4]

[1]Jožef Stefan Institute, Jamova 39, SI-1000 Ljubljana, Slovenia
[2]Department of Mechanical Eng., University of Western Macedonia, Kozani, GR-50100, Greece
[3]Iinst. of Material Science, NCSR Demokritos, Athens, GR-15310 Greece
[4]Inst. of Nuclear Technol. and Rad. Protection, NCSR Demokritos, Athens, GR-15310 Greece
[5]Hystore Technologies LtD, Ergates Industrial Area, 2643, Ergates, Nicosia, Cyprus
*corresponding author: tel.: +386 1 4773 552; fax: +386 1 25 19 385: e-mail: albert.prodan@ijs.si

**Abstract**

Titanium borates show promising hydrogen storage characteristics. Structural relaxation around individual hydrogen atoms and the binding energies are studied by means of the density functional theory methods for a number of hydrogenated *TiB$_2$*, *TiB* and *Ti$_2$B* structures. Starting with the possible symmetric hydrogen sites a random structure searching has been performed, in addition to locate all energetically stable adsorption sites. It is shown that for the three bulk compounds considered, the lowest binding energies are obtained for *TiB$_2$* (in the 0.3-1.8 eV range), the largest for *Ti$_2$B* (in the 3.9-4.7 eV range), while for *TiB* they are intermediate (in the 2.8-3.5 eV range). Calculations performed on hydrogenated *Ti$_2$B* result in two energetically stable sites for two different starting environments, suggesting a posible soft mode solution.

## 1. Introduction

Hydrogen represents with its high heating value a fuel of the future; it is environmentally friendly and regenerative. The main technological problem remains its storage and safe handling. Thus, searching for new storage methods and new materials represents a challenging research field[1-3]. A variety of hydrogen sorbents are currently considered. These include metal-organic[4-6] and covalent-organic frameworks[7,8], which show promising properties with regard to safe operation, fast kinetics and structural stability, but show at the same time a rather low adsorption energy for H$_2$ and consequently a relatively small storage capacity at room temperature and intermediate pressures[9,10]. Therefore, there is a need for new materials, whose H$_2$ adsorption energy would be higher, i.e., between 20–40 kJ/mole[11]. Higher H$_2$ binding energies were already achieved with sorbents decorated with transition-metals[12-28]. It was proposed that transition-metals added to fullerenes (C$_{60}$ or C$_{48}$B$_{12}$) and carbon nanotubes[13,16] represent hydrogen adsorbents with storage capacities as high as 8-9 wt%. Polymers decorated with transition-metals were also considered as promising hydrogen storage media[21-27]. Recently, a high weight percentage of H$_2$ uptake with rapid kinetics at room temperature was reported for transition-metal ethylene complexes, formed by laser ablation[28]. All these studies suggested that light metals like titanium, embedded into suitable skeletons made of other light elements, might represent ideal media for $H_2$ absorption. While titanium is characterized by a large $H_2$ binding energy, it also encounters a problem; its atoms, decorated on various nanostructures, tend to cluster[29]. Possible solutions include the usage of metals with weak metal–metal interactions or strong metal–supporter interactions[30], as well as possible enhancement of the metal–supporter interactions by means of incorporated defects[31] or by a direct integration of metal atoms into appropriate skeletons[32-35].

On the other hand boron, being lighter in comparison with carbon, represents a very good candidate to form titanium complexes with promising hydrogen storage characteristics[36-38]. Titanium-substituted closo-boranes[32] and tubular $TiB_2$ structures, based on graphite-like intercalated layered structures[33], were suggested as promising candidates[34]. It was shown that metal clustering can be avoided in calcium-doped boron fullerenes/nanotubes[39,40], where each calcium atom takes up as much as five $H_2$ molecules. In addition, several studies were performed on small boron clusters; it was shown that $B_x$ clusters with x up to 20 prefer planar ordering[41-45]. Such agglomerates also represent potential building blocks in complexes with polymers and transition-metals[46-49].

In the present work the hydrogenation as a function of the microstructural properties of $TiB_2$, $TiB$ and $Ti_2B$ is studied by means of the density functional theory (DFT) calculations, performed within the generalized-gradient-approximation functional and with the use of a plane-wave basis set. For each of the three compounds the possible hydrogen absorption sites are determined by considering the most likely interstitial positions with high coordination and, additionally, by ab-initio random structure searching studies[50], which give the minimum energy configurations for randomly chosen initial hydrogen sites. This approach is not only useful for finding absorption sites which are not expected on the basis of chemical intuition, but also for a rough estimation of the "basins of attraction" for the potential-energy-surface (PES) minima. Both can provide some information on the possible hydrogen diffusion pathways. Structural relaxation in the vicinity of the hydrogen absorption sites is studied by performing the calculations in enlarged super-cells.

In Sec. 2 we discuss the $TiB_x$ structures, provide the crystallographic data and list the a-priori absorption sites for the hydrogen atoms. In Sec. 3 the structural relaxation and the corresponding binding energies in the presence of hydrogen defects are presented. We conclude with a discussion on the relevance of our results for the practical aspects of hydrogen storage in the $TiB_x$ compounds.

## 2. The TiB$_x$ structures

### 2.1. The titanium di-boride TiB$_2$

$TiB_2$ crystallizes in the *AlB$_2$* or *ß-ThSi$_2$* structure type[51,52]. The structure is shown in *Fig.1* and the corresponding crystallographic data are given in *Table 1*. Four possible ideal H positions, given in the table with bold figures, are considered further. These positions are illustrated in *Fig.1b*.

*Table 1*

Crystallographic data of $TiB_2$ with four possible ideal H positions:
space group: P6/mmm (no. 191)
a = 0.3023 nm,  c = 0.3220 nm

| *atom* | *Wyckoff* | *x* | *y* | *z* |
|---|---|---|---|---|
| Ti | 1a | 0 | 0 | 0 |
| B | 2d | 1/3 | 2/3 | 1/2 |
| **H1** | **2c** | **2/3** | **1/3** | **0** |
| **H2** | **6i** | **0** | **1/2** | **1/4** |
| **H3** | **1b** | **0** | **0** | **1/2** |
| **H4** | **6k** | **4/5** | **0** | **1/2** |

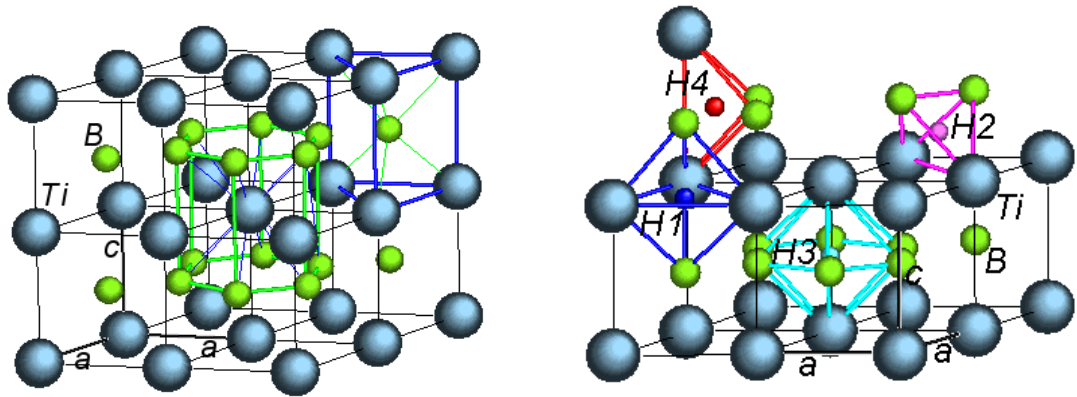

*Fig.1* The structure of *TiB$_2$* with the medium (green) balls representing B atoms in the trigonal prismatic coordination with Ti and the larger Ti atoms centered in 12-coordinated hexagonal prismatic coordination with B atoms (a). The small (dark blue, viollete, light blue and red) balls represent the four H ideal positions, listed in *Table 1* (b).

The ideal coordination of the H1 position is trigonal bipiramidal within two H-B distances of 0.161 nm and three H-Ti bonds of 0.175 nm. The B-H-B vertix is parallel to the *c*-axis, while the three Ti atoms form a triangle perpendicular to the *c*-axis with the H atom in the middle.
The H2 positions are surrounded by a deformed tetrahedron with two H-B bonds of 0.119 nm and two H-Ti bonds of 0.171 nm. The 2-fold axis of the tetrahedron is parallel to the *TiB$_2$ c*-axis. Six of such tetrahedra share faces with the H1 polyhedra.
The ideal H3 position is coordinated by six B atoms at 0.175 nm and two Ti atoms at 0.161 nm. The *TiB$_2$ c*-axis is parallel to the Ti-H-Ti vertix and perpendicular to the B hexagon.
The H4 positions are centered in deformed tetrahedra with two H-B bonds of 0.126 nm and two H-Ti bonds of 0.172 nm. Six of these tetrahedra compose into the enlaged H3 polyhedron.

*2.2.The titanium mono-boride TiB*

The mono-boride *TiB* crystallizes in the *FeB* structure type[51,53,54]. The structure is shown in *Fig.2* and the corresponding crystallographic data are given in *Table 2*. The structure is obtained from the one of *TiB$_2$* by removing from it every second trigonal prism. The remaining prisms form columns along the new orthorhombic *b*-direction. The structural tunnels thus formed collapse with the remaining trigonal prisms and change their coordination into single-capped trigonal prismatic. The symmetry is accordingly reduced from hexagonal into orthorhombic.
There are eight possible nonrelated H positions, three (H4, H7 and H8) in the more general (8d) and the remaining five (H1, H2, H3, H5 and H6) in (4c) positions. As shown below, four of these (H1, H3, H5 and H6) are energetically stable, while the others do not correspond to a PES minimum. Only the H1 positions are coordinated with four Ti atoms; these tetrahedra are edge-connected and form zig-zag chains along the orthorhombic *b*-direction. The remaining H positions are in 1B3Ti coordinations, except of H7 and H8, which occupy deformed 2B2Ti tetrahedra.

*Table 2*

Crystallographic data of the *TiB* structure with eight possible H positions; positions shown in bold are the energetically stable positions:
space group: Pnma (no. 62)
a = 0.6120 nm, b = 0.3060 nm, c = 0.4560 nm

| *atom* | *Wyckoff* | *x* | *y* | *z* |
|---|---|---|---|---|
| Ti | (4c) | 0.18000 | 1/4 | 1/8 |
| B | (4c) | 0.03600 | 1/4 | 0.61000 |
| **H1** | **(4c)** | **22/25** | **1/4** | **0.07300** |
| H2 | (4c) | 0.78600 | 3/4 | 0.50370 |
| **H3** | **(4c)** | **12/25** | **1/4** | **0.63000** |
| H4 | (8d) | 0.598 | 1/2 | 0.69125 |
| **H5** | **(4c)** | **0.286** | **3/4** | **0.24625** |
| **H6** | **(4c)** | **0.661** | **3/4** | **0.15625** |
| H7 | (8d) | 0.535 | 5/8 | 0.8475 |
| H8 | (8d) | 0.866 | 0 | 9/16 |

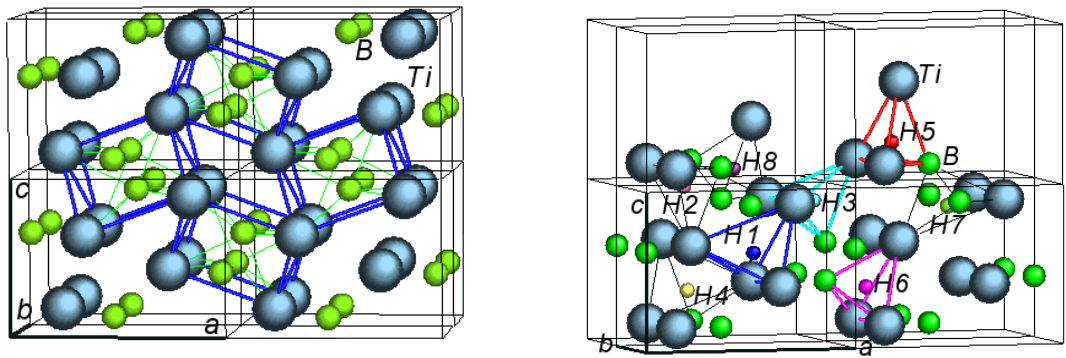

*Fig.2* The structure of *TiB* with the medium (green) and large (blue) balls representing B and Ti atoms (a). Small balls (of different colors) represent possible H positions, listed in *Table 2* with the corresponding bold (colored) tetrahedra representing the non-related stable interstices (b).

## 2.3. The titanium semi-boride Ti$_2$B

*Ti$_2$B* crystallizes in the $\theta$-*Al$_2$Cu* structure type[55-57]. The structure is shown in *Fig. 3a* and the relevant crystallographic data are collected in *Table 1*. By removing half of the tetragonal antiprismatic columns the structure type of *NbTe$_4$/TaTe$_4$* is obtained. From the three unrelated H positions, given in *Table 3* and shown in *Fig. 3b*, only two, i.e. H1 in 4Ti coordination and the two equivalent H3 sites in 3Ti1B coordination, were found to be energetically stable.

*Table 3*

Crystallographic data of the *Ti$_2$B* structure with the three symmetry unrelated ideal H positions; the two positions shown in bold are the energetically stable ones.
space group: I4/mcm (no.140)
a = 0.564 nm, c = 0.475 nm

| *atom* | *Wyckoff* | *x* | *y* | *z* |
|---|---|---|---|---|
| B | 4a | 0 | 0 | 1/4 |
| Ti | 8h | 1/6 | 1/3 | 0 |
| **H1** | **16l** | **1/8** | **5/8** | **13/16** |
| H2 | 4c | 0 | 0 | 0 |
| **H3a*** | **16l** | **5/12** | **17/24** | **11/16** |
| **H3b*** | **16l** | **1/12** | **5/24** | **5/16** |

*32 equivalent H3 sites are represented by two sets of symmetry-related 16l Wyckoff positions.

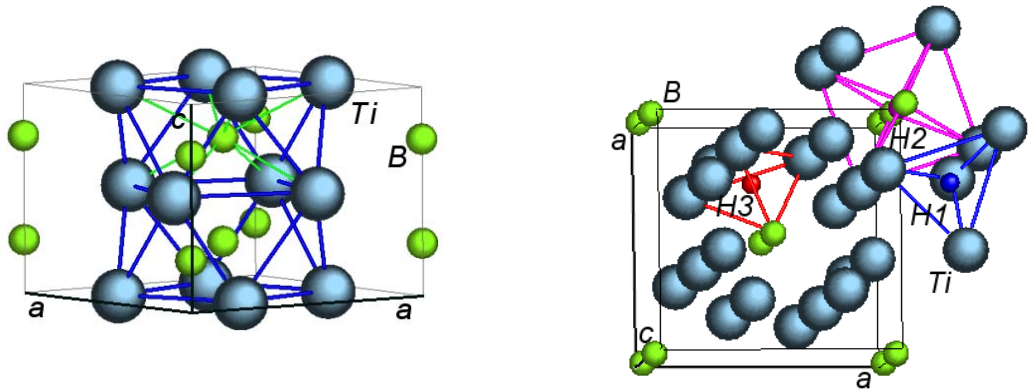

*Fig. 3* The crystal structure of Ti$_2$B with B atoms, represented by the middle sized (green) balls, in tetragonal anti-prismatic coordination with Ti atoms, represented by the large (blue) balls (a). Small (dark blue, violet and red) balls represent the hydrogen H1 (4Ti), H2 (4Ti2B) and H3 (3Ti1B) environments (b).

3. **The relaxation of the unit-cells and the atomic positions**

The calculations were performed using the plane-wave DFT code PWSCF (part of the Quantum Espresso package)[58] with the generalized gradient approximation (GGA) PBE functional[59]. Ultrasoft pseudopotentials with 400 Ry cutoff for the charge density and 40 Ry cutoff for the kinetic energy were used[60]. A 8x8x8 or 4x4x4 (dependent on the super-cell size) Monkhorst-Pack mesh of k-points was used with a cold-smearing by 0.02 Ry[61,62].

The experimental crystallographic data were used as the starting points for both, the relaxation of the unit cell and the relaxation of all atomic positions. This preliminary step was performed to test both, the experimental data and the numerical method used. The convergence threshold for the total energy was 0.5 meV and the threshold for forces was 10 meV/Å. The symmetry was not enforced during the minimization process, although in all cases considered no significant symmetry distortion took place.

Two series of calculations have been performed for each $TiB_x$ compound. In the first series the H atom was initially placed at one of the a-priori high-symmetry positions inside the cell, and then the whole structure was relaxed. For stable H positions, the relaxation of the H atom was always found to be relatively small. For unstable H positions, the H position was, however, displaced in the direction of the potential-energy-surface gradient until reaching the nearest local minimum. This first series of calculations already gave good indication about the possible H binding sites. Nevertheless, such a-priori initial positions based on chemical intuition are, in principle, subject to bias, since non-intuitive binding scenarios are not excluded. For this reason, we complemented this series of calculation by an additional one, where fully random (arbitrary) initial H positions were chosen. The relaxation calculations themselves were then performed in the same way as in the first series. This kind of studies are also known as the "ab-initio random structure searching". As detailed in the following, For $TiB_2$ and $TiB$ no new solutions were found, thus the value of performing the random structure searching was mainly in establishing the size of the basins of attraction for various local PES minima (the lowest energy solution, i.e., global minima, also tend to correspond to the solutions with the largest basins of attraction). For $Ti_2B$, however, random structure searching established the existence of pairs of solutions with similar energy, suggesting the presence of a soft mode in the system. The occurrence of this phenomenon would likely go undetected without the random structure search.

### 3.1. $TiB_2$

In accord with literature[51,52] the $TiB_2$ lattice parameters, obtained by GGA calculations, are a = 0.3026 nm and c = 0.3212 nm. To reduce the finite-size effects, calculations with H atoms in different environments have been performed in a ($2a$ x $2a$ x $2c$) super-cell with 8 Ti and 16 B atoms. The results are collected in *Table 4* and the relaxed environments of the possible four H interstices are shown in *Fig. 4*. The binding energy $E_{bind}$ is defined as the difference between the total energy of the $TiB_2$ cell containing one H atom and the sum of the energy of the pure $TiB_2$ cell and the energy of one H atom in empty space. By comparing the binding energies calculated for a ($2a$ x $2a$ x $2c$) super-cell with those for a ($1a$ x $1a$ x $1c$) cell, we found differences of the order of one percent, which was much less than the differences in $E_{bind}$ for different PES minima. The results indicate that the calculations may safely be performed in the eight times smaller cell and that the structural relaxation is small and of a very short range.

*Table 4*

Relaxed super-cell parameters, binding energies and atomic positions of either of the four H, eight Ti and sixteen B positions in the primitive ($2a$ x $2b$ x $2c$) $TiB_2$ super-cell. The Ti (0, 0, 0) position was kept in all cases fixed during the relaxation.

*unit cells:*

H1: a = 0.60774 nm, b = 0.60774 nm, c = 0.64267 nm, α = 89.9991°, β = 89.9999°, γ = 119.999°;
H2: a = 0.60728 nm, b = 0.60864 nm, c = 0.64491 nm, α = 90.0003°, β = 90.0002°, γ = 120.073°;
H3: a = 0.60671 nm, b = 0.60671 nm, c = 0.64756 nm, α = 90.0001°, β = 89.9998°, γ = 120.000°;
H4: a = 0.60734 nm, b = 0.60735 nm, c = 0.64664 nm, α = 89.9997°, β = 90.0005°, γ = 120.000°;

*binding energies*:
$E_{bind}$ (H1) = 1.79 eV; $E_{bind}$ (H2) = 1.28 eV; $E_{bind}$ (H3) = 0.26 eV; $E_{bind}$(H4) = 0.51 eV.

| *atom* | *starting positions* | | | *relaxed positions (H1)* | | | *relaxed positions (H2)* | | |
|---|---|---|---|---|---|---|---|---|---|
| | *x* | *y* | *z* | *x* | *y* | *z* | *x* | *y* | *z* |
| Ti | 0 | 0 | 0 | 0.0000 | 0.0000 | 0.0000 | 0.0000 | 0.0000 | 0.0000 |
| B | 1/6 | 1/3 | 1/4 | 0.1716 | 0.3383 | 0.2544 | 0.1751 | 0.3479 | 0.2650 |
| B | 1/3 | 1/6 | 1/4 | 0.3399 | 0.1700 | 0.2548 | 0.3335 | 0.1760 | 0.2548 |
| Ti | 0 | 0 | 1/2 | 0.0062 | 0.0031 | 0.5001 | 0.0000 | 0.0106 | 0.5041 |
| B | 1/6 | 1/3 | 3/4 | 0.1729 | 0.3370 | 0.7495 | 0.1670 | 0.3438 | 0.7522 |
| B | 1/3 | 1/6 | 3/4 | 0.3399 | 0.1700 | 0.7447 | 0.3335 | 0.1775 | 0.7539 |
| Ti | 0 | 1/2 | 0 | 0.0066 | 0.5033 | 0.0010 | 0.0000 | 0.5207 | 0.0000 |
| B | 1/6 | 5/6 | 1/4 | 0.1716 | 0.8334 | 0.2544 | 0.1668 | 0.8438 | 0.2549 |
| B | 1/3 | 2/3 | 1/4 | 0.3399 | 0.6700 | 0.2514 | 0.3335 | 0.6783 | 0.2548 |
| Ti | 0 | 1/2 | 1/2 | 0.0066 | 0.5033 | 0.5013 | 0.0000 | 0.5102 | 0.5041 |
| B | 1/6 | 5/6 | 3/4 | 0.1729 | 0.8360 | 0.7495 | 0.1678 | 0.8442 | 0.7515 |
| B | 1/3 | 2/3 | 3/4 | 0.3398 | 0.6699 | 0.7497 | 0.3335 | 0.6767 | 0.7539 |
| Ti | 1/2 | 0 | 0 | 0.5099 | 0.0000 | 0.0000 | 0.4989 | 0.0098 | 0.0036 |
| B | 2/3 | 1/3 | 1/4 | 0.6765 | 0.3383 | 0.2544 | 0.6665 | 0.3448 | 0.2548 |
| B | 5/6 | 1/6 | 1/4 | 0.8399 | 0.1700 | 0.2514 | 0.8249 | 0.1729 | 0.2650 |
| Ti | 1/2 | 0 | 1/2 | 0.5068 | 0.0031 | 0.5001 | 0.4975 | 0.0091 | 0.5058 |
| B | 2/3 | 1/3 | 3/4 | 0.6739 | 0.3369 | 0.7495 | 0.6665 | 0.3432 | 0.7539 |
| B | 5/6 | 1/6 | 3/4 | 0.8400 | 0.1701 | 0.7497 | 0.8330 | 0.1769 | 0.7522 |
| Ti | 1/2 | 1/2 | 0 | 0.5099 | 0.5099 | 0.0000 | 0.5011 | 0.5109 | 0.0036 |
| B | 2/3 | 5/6 | 1/4 | 0.6733 | 0.8366 | 0.2515 | 0.6665 | 0.8424 | 0.2548 |
| B | 5/6 | 2/3 | 1/4 | 0.8399 | 0.6700 | 0.2514 | 0.8332 | 0.6770 | 0.2549 |
| Ti | 1/2 | 1/2 | 1/2 | 0.5068 | 0.5037 | 0.5001 | 0.5025 | 0.5116 | 0.5058 |
| B | 2/3 | 5/6 | 3/4 | 0.6733 | 0.8366 | 0.7486 | 0.6665 | 0.8440 | 0.7539 |
| B | 5/6 | 2/3 | 3/4 | 0.8400 | 0.6699 | 0.7497 | 0.8322 | 0.6765 | 0.7515 |
| **H1** | 1/3 | 1/6 | 0 | **0.3398** | **0.1699** | **0.0310** | | | |
| **H2** | 0 | 1/4 | 1/8 | | | | **0.0000** | **0.2604** | **0.1033** |

| *atom* | *starting positions* | | | *relaxed positions (H3)* | | | *relaxed positions (H4)* | | |
|---|---|---|---|---|---|---|---|---|---|
| | *x* | *y* | *z* | *x* | *y* | *z* | *x* | *y* | *z* |
| Ti | 0 | 0 | 0 | 0.0000 | 0.0000 | 0.0000 | 0.0000 | 0.0000 | 0.0000 |
| B | 1/6 | 1/3 | 1/4 | 0.1685 | 0.3371 | 0.2606 | 0.1623 | 0.3332 | 0.2507 |
| B | 1/3 | 1/6 | 1/4 | 0.3371 | 0.1685 | 0.2606 | 0.3325 | 0.1768 | 0.2507 |
| Ti | 0 | 0 | 1/2 | 0.0000 | 0.0000 | 0.5212 | 0.0000 | 0.0000 | 0.5014 |
| B | 1/6 | 1/3 | 3/4 | 0.1675 | 0.3350 | 0.7606 | 0.1680 | 0.3333 | 0.7507 |
| B | 1/3 | 1/6 | 3/4 | 0.3350 | 0.1675 | 0.7606 | 0.3340 | 0.1668 | 0.7507 |
| Ti | 0 | 1/2 | 0 | 0.0000 | 0.5000 | 0.0112 | 0.0005 | 0.5005 | 0.0003 |
| B | 1/6 | 5/6 | 1/4 | 0.1685 | 0.8315 | 0.2606 | 0.1558 | 0.8233 | 0.2507 |
| B | 1/3 | 2/3 | 1/4 | 0.3333 | 0.6667 | 0.2606 | 0.3322 | 0.6618 | 0.2507 |
| Ti | 0 | 1/2 | 1/2 | 0.0000 | 0.5000 | 0.5100 | 0.0005 | 0.5005 | 0.5011 |
| B | 1/6 | 5/6 | 3/4 | 0.1675 | 0.8325 | 0.7606 | 0.1672 | 0.8332 | 0.7507 |
| B | 1/3 | 2/3 | 3/4 | 0.3333 | 0.6667 | 0.7606 | 0.3338 | 0.6656 | 0.7507 |
| Ti | 1/2 | 0 | 0 | 0.5000 | 0.0000 | 0.0112 | 0.5041 | 0.0000 | 0.0083 |
| B | 2/3 | 1/3 | 1/4 | 0.6667 | 0.3333 | 0.2606 | 0.6704 | 0.3382 | 0.2507 |
| B | 5/6 | 1/6 | 1/4 | 0.8315 | 0.1685 | 0.2606 | 0.8327 | 0.1666 | 0.2507 |
| Ti | 1/2 | 0 | 1/2 | 0.5000 | 0.0000 | 0.5100 | 0.5041 | 0.0000 | 0.5096 |
| B | 2/3 | 1/3 | 3/4 | 0.6667 | 0.3333 | 0.7606 | 0.6682 | 0.3344 | 0.7507 |
| B | 5/6 | 1/6 | 3/4 | 0.8325 | 0.1675 | 0.7606 | 0.8362 | 0.1677 | 0.7507 |
| Ti | 1/2 | 1/2 | 0 | 0.5000 | 0.5000 | 0.0112 | 0.5000 | 0.4995 | 0.0003 |
| B | 2/3 | 5/6 | 1/4 | 0.6629 | 0.8315 | 0.2606 | 0.6662 | 0.8334 | 0.2507 |
| B | 5/6 | 2/3 | 1/4 | 0.8315 | 0.6629 | 0.2606 | 0.8292 | 0.6668 | 0.2507 |
| Ti | 1/2 | 1/2 | 1/2 | 0.5000 | 0.5000 | 0.5100 | 0.5000 | 0.4995 | 0.5011 |
| B | 2/3 | 5/6 | 3/4 | 0.6650 | 0.8325 | 0.7606 | 0.6685 | 0.8324 | 0.7507 |
| B | 5/6 | 2/3 | 3/4 | 0.8325 | 0.6650 | 0.7606 | 0.8347 | 0.6668 | 0.7507 |
| **H3** | **0** | **0** | **1/4** | **0.0000** | **0.0000** | **0.2606** | | | |
| **H4** | **2/5** | **0** | **1/4** | | | | **0.4031** | **0.0001** | **0.2507** |

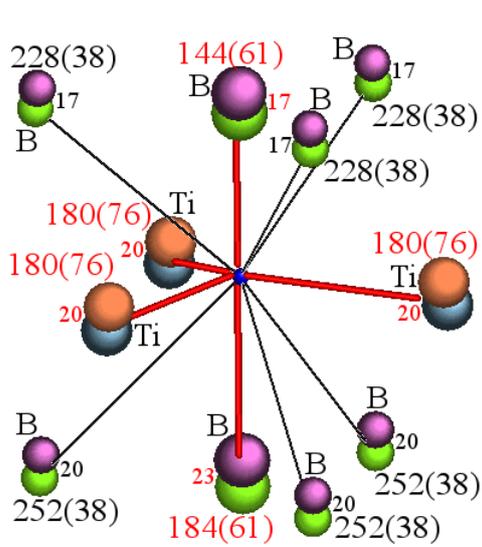

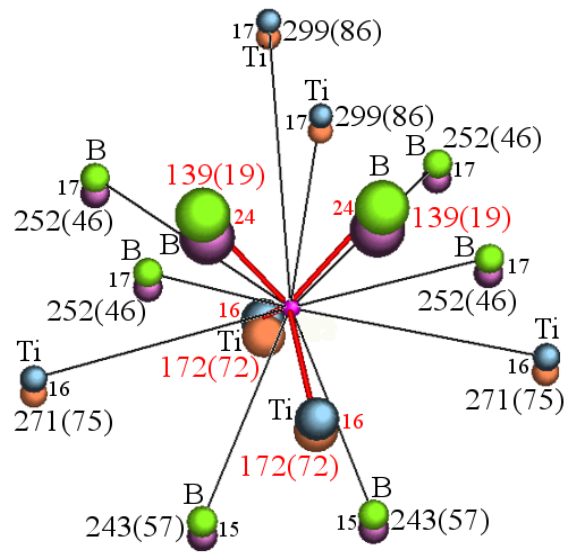

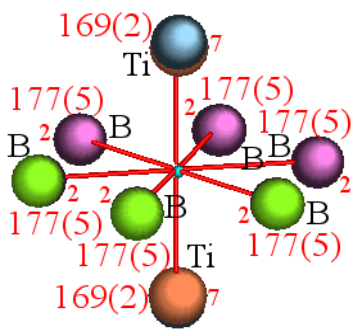

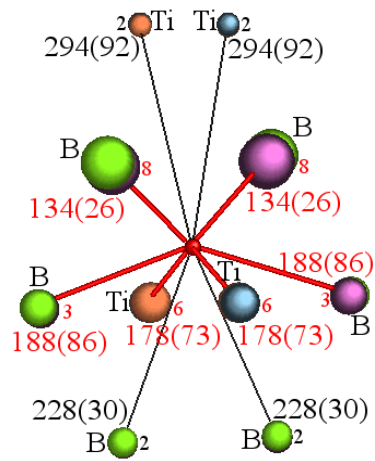

*Fig.4* The relaxed neighbourhood of the a) H1 (small and dark blue), b) H2 (small and violete), c) H3 (small and light blue) and d) H4 (small and red) atoms in the *TiB₂* structure. The sizes of the marked Ti (dark blue original and orange relaxed) and B (green original and violete relaxed) atoms are proportional to the reciprocal values of their distances from the H atom considered. The distances from the H atoms to the nearest displaced (bold and red) and second nearest displaced (light and black) atoms are given with large figures in $10^{-3}$ x nm (the last figures of the undisplaced positions in brackets). The corresponding small figures give the displacements in the same units. All atoms within 0.3 nm from the H atom are shown.

*3.2. TiB*

Structural relaxation within the GGA approach gives in case of pure *TiB* unit cell parameters (a = 0.610 nm, b = 0.304 nm and c = 0.455 nm) and atomic positions for Ti (0.1775 1/4 0.1225) and B (0.0302 1/4 0.59), which are in good agreement with the values from the literature[51,53,54], listed in *Table 2*. Calculations with an additional H atom in one of the four possible positions have been performed within the basic unit cell (1*a* x 1*b* x 1*c*) with four Ti and four B atoms. The results are collected in *Table 5* and the local environments of the H atoms are shown in *Fig. 5*.

*Table 5*

Relaxed unit cell parameters, binding energies and atomic positions of either of the four H, four Ti and four B positions within a primitive (1*a* x 1*b* x 1*c*) *TiB* unit cell. The Ti (0.18, 1/4, 1/8) position was in all cases kept fixed during the relaxation.

*unit cells*:
H1: a = 0.61031 nm, b = 0.30539 nm, c = 0.46103 nm, α = 90.000°, β = 89.573°, γ = 90.000°;
H3: a = 0.61577 nm, b = 0.30618 nm, c = 0.45844 nm, α = 90.000°, β = 90.411°, γ = 90.000°;
H5: a = 0.61238 nm, b = 0.30597 nm, c = 0.46161 nm, α = 90.000°, β = 89.616°, γ = 90.000°;
H6: a = 0.60855 nm, b = 0.30762 nm, c = 0.46151 nm, α = 90.000°, β = 90.205°, γ = 90.000°;

*binding energies*:
$E_{bind}$ (H1) = 3.54 eV; $E_{bind}$ (H3) = 3.20 eV; $E_{bind}$ (H5) = 2.98 eV; $E_{bind}$(H6) = 2.77 eV.

| *atom* | *starting positions* | | | *relaxed positions (H1)* | | | *relaxed positions (H3)* | | |
|---|---|---|---|---|---|---|---|---|---|
| | *x* | *y* | *z* | *x* | *y* | *z* | *x* | *y* | *z* |
| Ti | 0.18 | 1/4 | 1/8 | 0.18 | 1/4 | 1/8 | 0.18 | 1/4 | 1/8 |
| Ti | 17/25 | 1/4 | 3/8 | 0.6778 | 1/4 | 0.3792 | 0.6920 | 1/4 | 0.3691 |

| atom | starting positions | | | relaxed positions (H1) | | | relaxed positions (H3) | | |
|---|---|---|---|---|---|---|---|---|---|
| | x | y | z | x | y | z | x | y | z |
| Ti | 0.82 | 3/4 | 7/8 | 0.8206 | 3/4 | 0.8722 | 0.8306 | 3/4 | 0.8776 |
| Ti | 8/25 | 3/4 | 5/8 | 0.3257 | 3/4 | 0.6254 | 0.3187 | 3/4 | 0.6230 |
| B | 0.036 | 1/4 | 0.61 | 0.0329 | 1/4 | 0.5961 | 0.0324 | 1/4 | 0.6001 |
| B | 0.536 | 1/4 | 0.89 | 0.5248 | 1/4 | 0.9008 | 0.5331 | 1/4 | 0.9159 |
| B | 0.464 | 3/4 | 0.11 | 0.4713 | 3/4 | 0.0995 | 0.4745 | 3/4 | 0.0958 |
| B | 0.964 | 3/4 | 0.39 | 0.9767 | 3/4 | 0.4165 | 0.9752 | 3/4 | 0.4011 |
| **H1** | **22/25** | **1/4** | **0.073** | **0.8795** | **1/4** | **0.0722** | | | |
| **H3** | **12/25** | **1/4** | **0.63** | | | | **0.4733** | **1/4** | **0.6300** |

| atom | starting positions | | | relaxed positions (H5) | | | relaxed positions (H6) | | |
|---|---|---|---|---|---|---|---|---|---|
| | x | y | z | x | y | z | x | y | z |
| Ti | 0.18 | 1/4 | 1/8 | 0.18 | 1/4 | 1/8 | 0.18 | 1/4 | 1/8 |
| Ti | 17/25 | 1/4 | 3/8 | 0.6908 | 1/4 | 0.3869 | 0.6787 | 1/4 | 0.3887 |
| Ti | 0.82 | 3/4 | 7/8 | 0.8352 | 3/4 | 0.8830 | 0.8301 | 3/4 | 0.8711 |
| Ti | 8/25 | 3/4 | 5/8 | 0.3322 | 3/4 | 0.6387 | 0.3256 | 3/4 | 0.6264 |
| B | 0.036 | 1/4 | 0.61 | 0.0394 | 1/4 | 0.6044 | 0.0315 | 1/4 | 0.5964 |
| B | 0.536 | 1/4 | 0.89 | 0.5396 | 1/4 | 0.9090 | 0.5274 | 1/4 | 0.8936 |
| B | 0.464 | 3/4 | 0.11 | 0.4920 | 3/4 | 0.0953 | 0.4671 | 3/4 | 0.0924 |
| B | 0.964 | 3/4 | 0.39 | 0.9779 | 3/4 | 0.4126 | 0.9777 | 3/4 | 0.4039 |
| **H5** | **0.286** | **3/4** | **0.24625** | **0.3120** | **3/4** | **0.2472** | | | |
| **H6** | **0.661** | **3/4** | **0.15625** | | | | **0.6730** | **3/4** | **0.2072** |

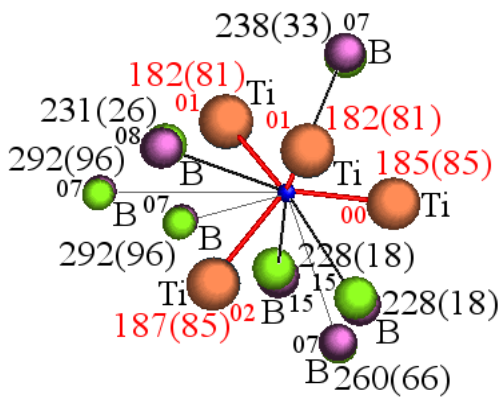
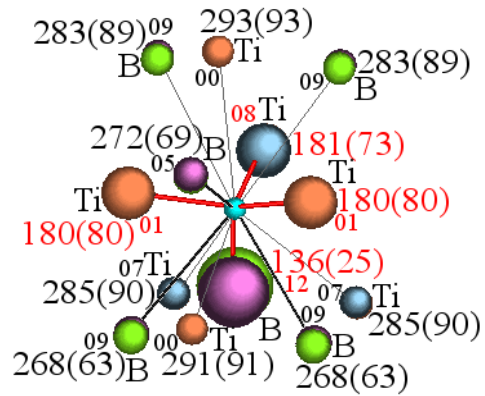
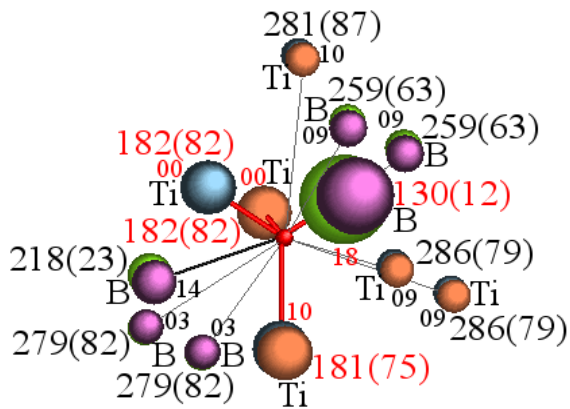
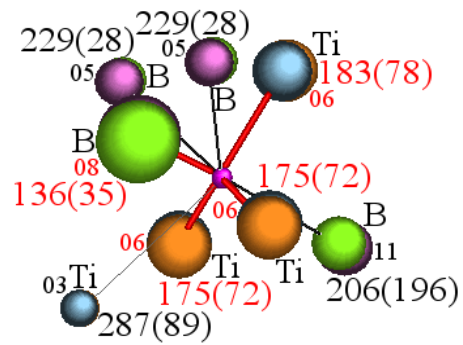

*Fig.5* The relaxed neighbourhood of the a) H1 (small and dark blue), b) H3 (small and light blue), c) H5 (small and red) and d) H6 (small and violet) atoms in the *TiB* structure. The sizes of the marked Ti (dark blue original and orange relaxed) and B (green original and violet relaxed) atoms are proportional to the reciprocal values of their distances from the H atoms considered. The distances from the H atoms to the nearest displaced (bold and red) and second nearest displaced (middle, light and black) atoms are given with large figures in $10^{-3}$ x nm (the last figures of the undisplaced positions in brackets). The corresponding small figures give the relaxed displacements in the same units. Shown are all atoms within 0.3 nm from the H atoms.

### 3.3. $Ti_2B$

The relaxed parameters (a = 0.564 nm and c = 0.475 nm), obtained by GGA calculations for the pure tetragonal body-centered Bravais lattice of $Ti_2B$ are again in accord with the data from the literature[55,56]. The same unit cell with eight Ti and four B atoms was then used in calculations with additional individual H atoms. The relaxed unit cell parameters and atomic positions, which correspond to the original values listed in *Table 3*, are given in *Table 6* together the corresponding binding energies. The original H2 position was found to be energetically unstable. Two slightly different stable arrangements were found for each of the remaining two H environment. These are shown in *Fig. 6* and marked as H1a and H1b for the H1 position in the tetrahedral 4Ti environment and as H3a and H3b for the H3 position in the 3Ti1B coordination.

*Table 6*

Relaxed positions of one H (H1a, H1b, H3a or H3b), eight Ti and four B positions in the body-centered $Ti_2B$ unit cell. The B (0, 0, 1/4) position was in all cases kept fixed during relaxation.

*unit cells*:
H1a: a = 0.5622 nm, b = 0.5665 nm, c = 0.4803 nm, α = 89.94°, β = 89.80°, γ = 88.82°;
H1b: a = 0.5709 nm, b = 0.5709 nm, c = 0.4738 nm, α = 90.04°, β = 90.04°, γ = 90.22°;
H3a: a = 0.5669 nm, b = 0.5632 nm, c = 0.4795 nm, α = 89.81°, β = 90.47°, γ = 89.69°;
H3b: a = 0.5724 nm, b = 0.5691 nm, c = 0.4733 nm, α = 90.01°, β = 90.23°, γ = 90.28°;

*binding energies*:
$E_{bind}$ (H1a) = 4.66 eV, $E_{bind}$ (H1b) = 4.16 eV, $E_{bind}$ (H3a) = 4.25 eV and $E_{bind}$ (H3b) = 3.88 eV.

| atom | starting positions | | | relaxed positions (H1a) | | | relaxed positions (H1b) | | |
|---|---|---|---|---|---|---|---|---|---|
| | *x* | *y* | *z* | *x* | *y* | *z* | *x* | *y* | *z* |
| B | 0 | 0 | 1/4 | 0 | 0 | 1/4 | 0 | 0 | 1/4 |
| B | 1/2 | 1/2 | 1/4 | 0.4947 | 0.4931 | 0.1189 | 0.4939 | 0.5069 | 0.2502 |
| Ti | 1/6 | 1/3 | 0 | 0.1511 | 0.3186 | 0.9344 | 0.1696 | 0.3385 | 0.0530 |
| B | 0 | 0 | 3/4 | 0.9988 | 0.9935 | 0.6214 | 0.9976 | 0.0068 | 0.8556 |
| B | 1/2 | 1/2 | 3/4 | 0.5003 | 0.4963 | 0.7515 | 0.5004 | 0.5040 | 0.8557 |
| Ti | 2/3 | 1/6 | 0 | 0.6757 | 0.1466 | 0.9373 | 0.6575 | 0.1640 | 0.0485 |
| Ti | 5/6 | 2/3 | 0 | 0.8456 | 0.6698 | 0.9341 | 0.8322 | 0.6759 | 0.0528 |
| Ti | 1/3 | 5/6 | 0 | 0.3220 | 0.8433 | 0.9359 | 0.3342 | 0.8408 | 0.0510 |
| Ti | 5/6 | 1/3 | 1/2 | 0.8147 | 0.3339 | 0.4386 | 0.8315 | 0.3378 | 0.5526 |
| Ti | 1/3 | 1/6 | 1/2 | 0.3410 | 0.1718 | 0.4331 | 0.3311 | 0.1725 | 0.5543 |

| atom | | | | | | | | | |
|------|---|---|---|---|---|---|---|---|---|
| Ti | 1/6 | 2/3 | 1/2 | 0.1880 | 0.6494 | 0.4380 | 0.1666 | 0.6730 | 0.5496 |
| Ti | 2/3 | 5/6 | 1/2 | 0.6516 | 0.8259 | 0.4408 | 0.6662 | 0.8375 | 0.5541 |
| **H1a** | **7/8** | **5/8** | **5/16** | **0.8862** | **0.6192** | **0.3097** | | | |
| **H1b** | **7/8** | **3/8** | **3/16** | | | | **0.8754** | **0.3816** | **0.1753** |

| *atom* | *starting positions* | | | *relaxed positions (H3a)* | | | *relaxed positions (H3b)* | | |
|--------|---|---|---|---|---|---|---|---|---|
| | *x* | *y* | *z* | *x* | *y* | *z* | *x* | *y* | *z* |
| B  | 0    | 0    | 1/4  | 0      | 0      | 1/4    | 0      | 0      | 1/4    |
| B  | 1/2  | 1/2  | 1/4  | 0.4960 | 0.4990 | 0.1215 | 0.5000 | 0.5042 | 0.2495 |
| Ti | 1/6  | 1/3  | 0    | 0.1495 | 0.3185 | 0.9364 | 0.1628 | 0.3426 | 0.0538 |
| B  | 0    | 0    | 3/4  | 0.9994 | 0.9960 | 0.6214 | 0.9995 | 0.0114 | 0.8628 |
| B  | 1/2  | 1/2  | 3/4  | 0.4954 | 0.4882 | 0.7535 | 0.5020 | 0.5045 | 0.8589 |
| Ti | 2/3  | 1/6  | 0    | 0.6759 | 0.1486 | 0.9356 | 0.6616 | 0.1669 | 0.0570 |
| Ti | 5/6  | 2/3  | 0    | 0.8468 | 0.6721 | 0.9350 | 0.8384 | 0.6587 | 0.0648 |
| Ti | 1/3  | 5/6  | 0    | 0.3159 | 0.8490 | 0.9501 | 0.3442 | 0.8451 | 0.0547 |
| Ti | 5/6  | 1/3  | 1/2  | 0.8195 | 0.3427 | 0.4365 | 0.8308 | 0.3357 | 0.5558 |
| Ti | 1/3  | 1/6  | 1/2  | 0.3440 | 0.1714 | 0.4373 | 0.3320 | 0.1726 | 0.5530 |
| Ti | 1/6  | 2/3  | 1/2  | 0.1656 | 0.6471 | 0.4270 | 0.1825 | 0.6646 | 0.5432 |
| Ti | 2/3  | 5/6  | 1/2  | 0.6615 | 0.8241 | 0.4223 | 0.6604 | 0.8363 | 0.5444 |
| **H3a** | **5/12** | **17/24** | **11/16** | **0.4297** | **0.7008** | **0.6440** | | | |
| **H3b** | **11/12** | **19/24** | **11/16** | | | | **0.9408** | **0.8040** | **0.7451** |

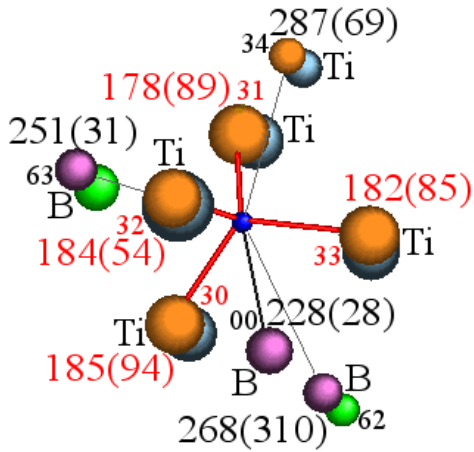
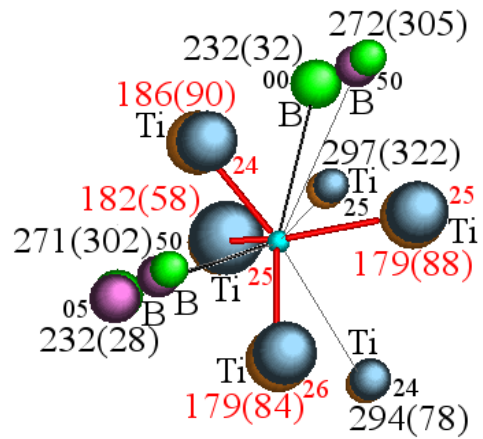
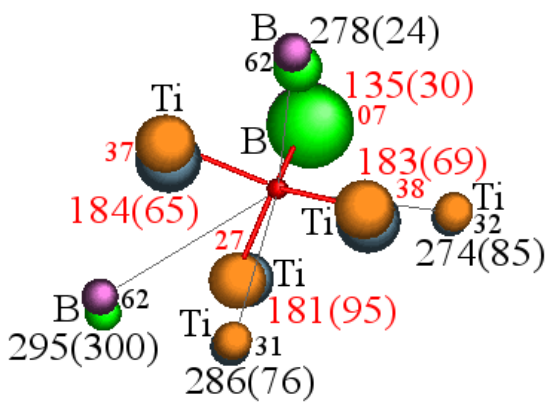
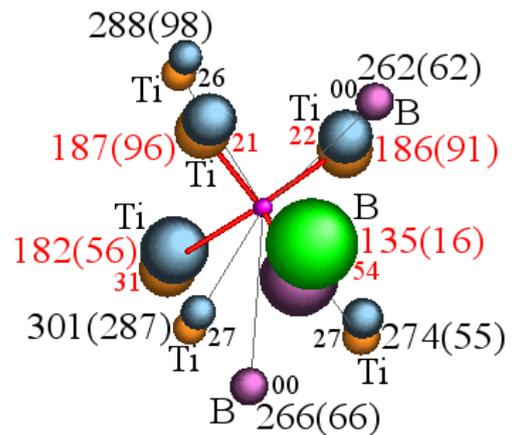

*Fig.6* The relaxed neighbourhoods of the a) H1a (small and dark blue), b) H1b (small and light blue), c) H3a (small and red) and d) H3b (small and violete) atomic positions in the *Ti$_2$B* structure. The sizes of the Ti (dark blue original and orange relaxed) and B (green original and violete relaxed) atoms are proportional to the reciprocal values of their distances from the H atoms. The distances from the H atoms to the nearest displaced (bold and red) and second nearest displaced (middle, light and black) atoms are given with large figures in $10^{-3}$ x nm (the last figures of the undisplaced positions in brackets). The corresponding small figures show the displacements in the same units. All atoms within a sphere with a radius slightly larger than 0.3 nm from the H atoms are shown.

## 4. Discussion

It was shown recently[34] that metal diborides, particularly in the form of nanotubes, represent very promising H storage media with molecular H$_2$ binding energies in the range of 0.2 to 0.6 eV/H$_2$ and a very high theoretical retrievable H$_2$ storage density (of the order of almost 8 wt %). The binding energies of individual H atoms, obtained in our calculations, are in range of 0.26 eV to 1.78 eV for *TiB$_2$*, while the binding energies obtained in case of the other two hydrogenated compounds, i.e. the mono- and semi-borides, appear much higher; for the four hydrogenated Ti mono-borides the H binding energies fall between 2.77 eV and 3.54 eV and for the semi-borides even higher, between 3.88 eV and 4.66 eV. However, direct comparisons of these values with regard to possible hydrogen storage are not possible. First of all, as hydrogen storage media compounds are desired, able to absorb larger amounts of hydrogen, while in our calculations single H atoms and their environments were considered. Also, the energy obtained for a H$_2$ molecule has to be replaced in case of a free H atom by the dissociative energy of the molecule in vacuum (4.48 eV)[63]. It was shown recently[64] that relevant for hydrogen storage are compounds with changes in enthalpy between 30 and 70 kJ/mole H$_2$, which corresponds to H binding energies between 2.4 and 2.7 eV/H atom. These values would roughly correspond to those obtained for the hydrogenated *TiB* compounds.

The structural adjustment required by the presence of the H atom was found to be in most cases small, with only the first neighbours displaced appreciably. However, a certain minor influence of the next, periodically positioned H atom onto the next-nearest metal atoms cannot be completely excluded. To show that this influence is small, the initial calculations (on the *TiB$_2$*-based structures) were performed in enlarged super-cells. The remaining calculations (on *TiB*- and *Ti$_2$B*-based structures) were performed with the basic unit cells only; this made it possible to perform random structure search calculations with more initial starting positions for the H atom. The random structure search was performed for a few tens of individual positions for each compound considered. It is thus very likely that we have located all PES minima. We also observe that the H adsorption sites with a strong binding correspond to large bassins of attraction in the position space.

Typical ionic hydrides are binary alkali-metal hydrides (e.g. LiH with the sodium chloride type of structure)[65] and alkaline-earth metal hydrides (e.g. MgH$_2$ with the rutile type of structure)[66], while most other complex metal hydrides[67] are characterized by mixed ionic/covalent bonding. A rather promising example with regard to H absorption/desorption kinetics is potassium aluminum hydride KAlH$_4$ with Al-H distances of about 0.165 nm[68]. Contrary, interstitial hydrides are characterized by metallic bonding. A well known example is Pd, whose face-centred cubic lattice parameter of 0.38874 nm is changed only slightly during hydrogenation, with Pd-H distances comparable to the Al-H distances in $KAlH_4$. The shortest Ti-H bonds in the three $TiB_x$ compounds considered range between 0.170 and 0.185 nm, while the corresponding B-H distances are shorter, only between 0.130 and 0.135 nm. These compounds belong to the interstitial hydrides. They are lately intensively studied, because they show good hydrogenation and kinetic properties, particularly in the form of nanotubes and other low-dimensional forms[34,69].

### *4.1. TiB$_2$*

As a result of the lattice relaxation around the H1 atom one of two B-H bonds is enlarged, while the second is shortened. Thus, the starting 3Ti2B H position is relaxed into an almost regular 3Ti1B tetrahedron with all three Ti-H and one B-H distances of about 0.18 nm. The remaining B-H bond is shorter (only 0.14 nm) and forms with the other four nearest neighbors a single-capped tetrahedron around the H1 atom. The next-nearest neighbors of H1 are only slightly accommodated to fit the relaxation.

The starting 2Ti2B coordination of the H2 atom is relaxed by enlarging the two nearest B-H distances symmetrically from 0.12 nm to 0.14 nm, while the nearest Ti-H bonds remain practically unchanged at 0.17 nm. Again, all remaining next-nearest Ti and B atoms are accordingly accommodated.

The H3 and H4 interstitial positions require only minor accommodation of the original *TiB$_2$* structure. The relaxation of the H3 position hardly influences its starting 2Ti6B environment. Likewise, in case of H4 the rather deformed starting tetrahedraly-centered 2Ti2B coordination undergoes only a minor enlargement. The largest, but still relatively small displacements involve both nearest neighbor B-H bonds (0.13 nm), while the two nearest Ti-H bonds (0.18 nm) remain after relaxation only slightly shorter from the next-nearest B-H pairs (0.19 nm). Thus, the calculations clearly reveal that H3 and H4 are energetically stable positions within their respective original environments.

### *4.2. TiB*

In case of the H1 atom, the four Ti-H bonds of the starting 4Ti tetragonal coordination undergo with relaxation only minor displacements. The calculations performed within the (1*a* x 1*b* x 1*c*) basic structure unit cell show that the next-nearest Ti and B positions are practically not affected by the relaxation.

The main relaxation around the H3 position involves a slight elongation of the shortest B-H bond (0.14 nm), while the three Ti-H bonds of the 3Ti1B tetrahedron remain practically unchanged (0.18 nm). Again, the second-nearest neighbors are hardly affected by the relaxation.

A very similar situation to the one of H3 takes place in case of both, the H5 and H6 atoms. Although slightly enlarged, the B-H bond of the 3Ti1B tetrahedron remains short (0.13 nm) in comparison with the remaining three Ti-H bonds (0.18 nm).

### *4.3.Ti$_2$B*

In case of H1a and H1b the H atom is centered in an almost regular tetrahedral coordination with 4Ti atoms at the corners. As a result of relaxation, the four Ti-H bonds are in both cases only slightly modified. In case of H3a and H3b the two relaxations give again very similar results; both leave the three Ti-H bonds practically unchanged (0.18 nm), while the remaining B-H bond is only slightly enlarged (0.14 nm) and leaves the relaxed 3Ti1B tetrahedron deformed. In general, the displacements being the result of the relaxation, they appear somewhat larger than in *TiB$_2$* and *TiB*. Thus, only the positions of the nearest neighbors to the H atom should be taken as reliable, while the next-nearest Ti and B atoms seem to be appreciably influenced by the next-nearest H atoms. The two pairs of stable H positions correspond to two sets of relaxed unit-cell parameters; those obtained for H1a and H3a are similar and so are the parameters obtained for H1b and H3b atoms. This may suggest a soft-mode type of deformation with two stable solutions.

## 5. Conclusions

The following general conclusions follow from our numerical studies:
- The calculations show that the lattice relaxation caused by the presence of single H atoms in various $TiB_2$, $TiB$ and $Ti_2B$ interstitial sites involves appreciably only the nearest environments of the H atoms.
- The binding energies of the H atoms are the lowest for the stable $TiB_2$ sites (between 0.26 eV and 1.78 eV), somewhat larger for the hydrogenated $TiB$ compounds (between 2.77 eV and 3.54 eV) and the largest in case of $Ti_2B$ (between 3.88 eV and 4.66 eV).
- Although direct comparisons with other calculations are applicable, the lattice relaxation and the corresponding binding energies obtained suggest that the Ti di-, mono- and semi-borides are promising candidates for hydrogen storage applications.
- The energetically stable coordination polyhedra around H sites include rather symmetrical 4Ti or 3Ti1B tetrahedra (the second with an additional single-capped shorter B-H bond), deformed 3Ti1B tetrahedral sites (with the only B-H bond appreciably shorter from the remaining three Ti-B bond), and very deformed 2Ti2B tetrahedra (with the pair of B-H bond much shorter from the remaining Ti-H pair).
- Two energetically stable sites were found for one environment in case of $TiB_2$ and for two such environments in case of $Ti_2B$. In the second case the four H positions with the corresponding two relaxed unit cells suggest a possible soft-mode solution.

## Acknowledgments

Financial support of the Slovenian Research Agency (ARRS) (RŽ, HJPvM, EZ and AP), the bilateral cooperation program between the Hellenic Republic and the Republic of Slovenia (GSRT Code 043Γ) and the ATLAS-H2 European Project PIAP-GA-2009-251562 is gratefully acknowledged.